\documentclass[twoside,twocolumn,9pt]{article}
\usepackage{extsizes}
\usepackage[super,sort&compress,comma]{natbib} 
\usepackage[version=3]{mhchem}
\usepackage[left=1.5cm, right=1.5cm, top=1.785cm, bottom=2.0cm]{geometry}
\usepackage{balance}
\usepackage{times,mathptmx}
\usepackage{sectsty}
\usepackage{graphicx} 
\usepackage{lastpage}
\usepackage[format=plain,justification=justified,singlelinecheck=false,font={stretch=1.125,small,sf},labelfont=bf,labelsep=space]{caption}
\usepackage{float}
\usepackage{fancyhdr}
\usepackage{fnpos}
\usepackage[english]{babel}
\usepackage{array}
\usepackage{droidsans}
\usepackage{charter}
\usepackage[T1]{fontenc}
\usepackage[usenames,dvipsnames]{xcolor}
\usepackage{setspace}
\usepackage[compact]{titlesec}

\usepackage{epstopdf}

\definecolor{cream}{RGB}{222,217,201}

\begin{document}

\pagestyle{fancy}
\thispagestyle{plain}
\fancypagestyle{plain}{

}

\makeFNbottom
\makeatletter
\renewcommand\LARGE{\@setfontsize\LARGE{15pt}{17}}
\renewcommand\Large{\@setfontsize\Large{12pt}{14}}
\renewcommand\large{\@setfontsize\large{10pt}{12}}
\renewcommand\footnotesize{\@setfontsize\footnotesize{7pt}{10}}
\makeatother

\renewcommand{\thefootnote}{\fnsymbol{footnote}}
\renewcommand\footnoterule{\vspace*{1pt}%
\color{cream}\hrule width 3.5in height 0.4pt \color{black}\vspace*{5pt}} 
\setcounter{secnumdepth}{5}

\makeatletter 
\renewcommand\@biblabel[1]{#1}            
\renewcommand\@makefntext[1]%
{\noindent\makebox[0pt][r]{\@thefnmark\,}#1}
\makeatother 
\renewcommand{\figurename}{\small{Fig.}~}
\sectionfont{\sffamily\Large}
\subsectionfont{\normalsize}
\subsubsectionfont{\bf}
\setstretch{1.125} 
\setlength{\skip\footins}{0.8cm}
\setlength{\footnotesep}{0.25cm}
\setlength{\jot}{10pt}
\titlespacing*{\section}{0pt}{4pt}{4pt}
\titlespacing*{\subsection}{0pt}{15pt}{1pt}

\fancyfoot{}
\fancyfoot[RO]{\footnotesize{\sffamily{1--\pageref{LastPage} ~\textbar  \hspace{2pt}\thepage}}}
\fancyfoot[LE]{\footnotesize{\sffamily{\thepage~\textbar\hspace{3.45cm} 1--\pageref{LastPage}}}}
\fancyhead{}
\renewcommand{\headrulewidth}{0pt} 
\renewcommand{\footrulewidth}{0pt}
\setlength{\arrayrulewidth}{1pt}
\setlength{\columnsep}{6.5mm}
\setlength\bibsep{1pt}

\makeatletter 
\newlength{\figrulesep} 
\setlength{\figrulesep}{0.5\textfloatsep} 

\newcommand{\topfigrule}{\vspace*{-1pt}%
\noindent{\color{cream}\rule[-\figrulesep]{\columnwidth}{1.5pt}} }

\newcommand{\botfigrule}{\vspace*{-2pt}%
\noindent{\color{cream}\rule[\figrulesep]{\columnwidth}{1.5pt}} }

\newcommand{\dblfigrule}{\vspace*{-1pt}%
\noindent{\color{cream}\rule[-\figrulesep]{\textwidth}{1.5pt}} }

\makeatother

\twocolumn[
  \begin{@twocolumnfalse}
\vspace{3cm}
\sffamily
\begin{tabular}{m{4.5cm} p{13.5cm} }

& \noindent\LARGE{\textbf{Microscopic Details of a Fluid/Thin Film Triple Line$^\dag$}} \\
%
%
\vspace{0.3cm} & \vspace{0.3cm} \\

 & \noindent\large{Timothy Twohig,\textit{$^{a}$} Sylvio May\textit{$^{a}$} and Andrew B. Croll,$^{\ast}$\textit{$^{a}$}} \\
\\
& \noindent\normalsize{In recent years, there has been a considerable interest in the mechanics of soft objects meeting fluid interfaces (elasto-capillary interactions).  In this work we experimentally examine the case of a fluid resting on a thin film of rigid material which, in turn, is resting on a fluid substrate.  To simplify complexity, we adapt the experiment to a one-dimensional geometry and examine the behaviour of polystyrene and polycarbonate films directly with confocal microscopy.  We find that the fluid meets the film in a manner consistent with the Young-Dupr\'{e} equation when the film is thick, but transitions to what appears similar to a Neumann like balance when the thickness is decreased.  However, on closer investigation we find that the true contact angle is always given by the Young construction.  The apparent paradox is a result of macroscopically measured angles not being directly related to true microscopic contact angles when curvature is present.  We model the effect with the Euler-Bernoulli beam on a Winkler foundation as well as with an equivalent energy based capillary model.  Notably, the models highlight several important lengthscales and the complex interplay of tension, gravity and bending in the problem.
} \\

\end{tabular}

 \end{@twocolumnfalse} \vspace{0.6cm}

  ]

\renewcommand*\rmdefault{bch}\normalfont\upshape
\rmfamily
\section*{}
\vspace{-1cm}


\footnotetext{\textit{$^{a}$~Department of Physics, North Dakota State University, Fargo, USA.  Tel: +1 413 320 3810; E-mail: andrew.croll@ndsu.edu}}

\footnotetext{\dag~Electronic Supplementary Information (ESI) available: [details of any supplementary information available should be included here]. See DOI: 10.1039/cXsm00000x/}




\section{Introduction}
The interplay between capillary interactions and deformable materials has recently drawn great interest from the soft matter community.  The focus partly stems from the perceived utility of capillary forces in guiding the assembly of microstructures\cite{Terfort1997} or even in guiding the folding of origami inspired devices.\cite{py2007,py2009}  Interest has also grown from the identification of novel phenomena which occur simply because capillary forces are applied to solids compliant enough to be locally distorted.\cite{Pericet2008, Shanahan1995, Extrand1996, Jerison2011, Style2012, Style2013b,Qin2018, Pham2017, Style2013a, Hui2002, Mora2010, Andreotti2016, Style2017}  A simple example can be seen when a fluid drop is placed on a flat elastic half-space.  If the modulus of the elastic substrate is large, the drop adopts the familiar a spherical cap shape with a contact angle, $\theta_{Y}$, determined by a balance of surface energies in the horizontal direction.  However, if the modulus of the substrate is decreased, the substrate is pulled by the vertical component of the contact line into a cusp shape, reminiscent of the Neumann construction.\cite{Jerison2011, Style2012, Style2013b}

Decreasing Young's modulus is not the only way to increase deformability; a slender material will also be easily deformed as its thickness is reduced.  For example, a drop added to a thin film may lead to bending on the scale of the sheet,\cite{py2007,py2009,Bae2015} interesting pattern formation,\cite{Huang2007,Schroll2013,Paulsen2016,Paulsen2017} or even complete droplet wrapping\cite{Paulsen2015,Kumar2018}.  The critical difference between slender and soft deformations lies in the fact that slender, high-modulus films strongly resist in plane stretching.  The significant resistance to creation of Gaussian curvature in a thin rigid sheet, creates new complexity in understanding how the curved fluid-solid triple line of a droplet meets the orthogonally curved sheet.  In this work, we examine the interaction of a fluid with a thin rigid sheet on the microscopic level.  We work in one dimension in order to avoid in plane stretching, buckling and localization.\cite{Paulsen2017}  We find that the apparent contact triple line changes continuously from solid-like (Young-Dupr\'{e}) to fluid-like (Neumann) as the film's thickness is reduced.  However, we note that true contact is always Young-like, and force balance arguments constructed from macroscopically observed angles are not always correct.  Importantly, we show how film bending, gravity of the fluid substrate and external tension contribute to the deformation shape and apparent contact angles.

Fluid drops on thin, solid substrates have been studied macroscopically by several researchers, though little effort has been made to reconcile macrocsopic models with the microscale.  Russell and co-workers were the first to explore the effect of a fluid drop resting on a film floating on a water bath.\cite{Huang2007}  The capillary action of a droplet draws the film in towards the droplet, creating a hoop stress which quickly leads to the buckling of the film.  The buckling can be modeled through the application of the F\"{o}ppl-von K\'{a}rm\'{a}n equations in a limit where bending is ignored.\cite{Schroll2013}  The buckling, while useful in determining the film's modulus, significantly complicates any effort to observe the film/fluid triple-line where film bending is important.  This is especially true if the film is pushed to localization.\cite{Paulsen2017}  

On the other hand, if a droplet is placed on a thin film which has its boundaries fixed no buckling will occur on microscopic scales.\cite{Schulman2015,Schulman2017,Schulman2018,Nadermann2013,Hui2014}  In this case a drop deforms the film underneath it into a parabolic shape (macroscopically), which can be modeled again with a simplified F\"{o}ppl-von K\'{a}rm\'{a}n equation neglecting bending and any stretching in excess of what is created by the clamped boundaries.\cite{Schulman2017}  Macroscopically this view is quite successful, leading to direct measurements of local strain variation\cite{Nadermann2013,Schulman2015} and strain dependent surface energies in polymer glasses\cite{Schulman2018} among other results. However, there is no clear picture of what happens in the film at small lengthscales where again stretching and bending may be important.

\begin{figure}[h]
\centering
  \includegraphics[ trim={30 00 00 0},clip,width = 0.48\textwidth ]{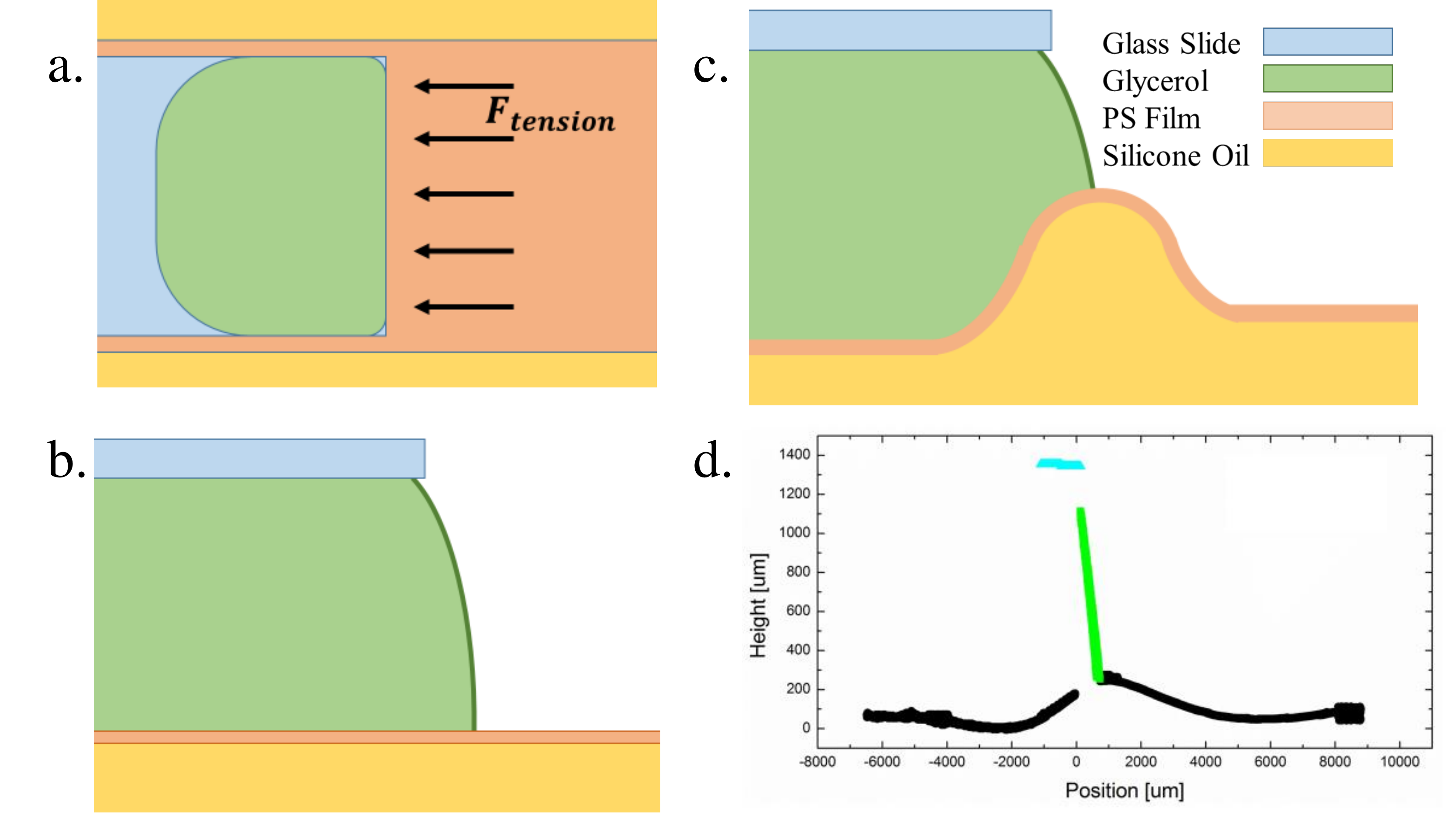}
  \caption{Schematic of the experimental setup. a.) Top view of the glycerol drop on polymer film floating on oil bath. Glycerol edge is straightened by the glass slide reducing the problem to one-dimension. b.) Side view of the initial setup of the experiment. c.) Side view of the final state of the experiment. d.) Profile data of a real film extracted from the three-dimensional confocal microscope scan of a 4.3 micron thick PC film.  Note the oscillatory profile is strong evidence that film bending cannot be ignored.}
  \label{Expt}
\end{figure}

In this paper we describe a simple, one-dimensional experiment which significantly reduces complexity and allows a direct observation of the local shape of a thin film in contact with a fluid.  In short, a glassy polymer film (polystyrene, PS, or polycarbonate, PC) floating on a fluid surface is placed under a droplet of a second fluid which is deformed into a long straight contact line by a capping glass slide (see Fig.~\ref{Expt}).  The film surfaces are located in three dimensions through laser scanning confocal microscopy and contact angles are determined on a microscopic scale.  Varying the thickness of the film reveals that the deformation persists even in films on the order of 10's of microns in thickness ( F\"{o}ppl-von K\'{a}rm\'{a}n number $\gamma \sim L^2/t^2 \sim 10^{4}$, where $L$ is the lateral size of the film and $t$ is the film thickness).  As the film thickness drops below $\sim 1$ micron we find the contact becomes altered, reminiscent of the Neumann construction typical of fluid/fluid contact problems.  However, quantitatively the Neumann result does not appear to be valid in the range of thicknesses we examine.

The contact can be modeled with an analytic capillary + bending model, or equivalently via the Euler-Bernoulli beam coupled to a Winkler foundation. The models aid in the discussion of the experimental results, and offer new insight into related problems.  We find the data to be well represented by these simple theories, and we find the theory is consistant with a Young-Dupr\'{e} force balance occurring at the triple line.  This occurs because the film is locally flat at the point of contact (the maximum of the film contour).  The film still rises accounting for the unbalanced vertical component of the contact line, but the results are not simply calculated from surface forces alone.  Understanding film shape requires knowledge of the relationship of bending, fluid weight, and film tension.  The models highlight several important lengthscales which can be used to qualify dominant features in the problem, which we discuss in relation to several limiting cases (for example tension free films, zero density fluid substrates) which are inaccessible to experiments.


\section{Experimental}
\subsection{Film Preparation}
Solutions of polystyrene (PS) were created by dissolving bulk PS (Aldrich, Mw = 192~kg/mol) in toluene.  To facilitate imaging, Nile Red fluorescent dye (MP Biomedicals, LLC) was added to the solution.  The solutions were between 0.1\% and 10\% PS by weight, depending on the desired film thickness.  Polycarbonate (PC) films were created in a similar manner by dissolving bulk PC (Scientific Polymer Products Inc., Mw = 60~kg/mol) in chloroform (Nile Red was also added to these solutions).  PC solutions were made between 0.1\% and 10\% by weight. Thin films were created by spin coating or drop casting the PS or PC solutions onto freshly cleaved mica sheets (Ted Pella, Inc.). Spin coating was used to create polymer samples of thickness 10~nm-700~nm, and drop casting was used to create polymer films of thickness 0.700~$\mu$m - 10~$\mu$m. Samples were annealed at a temperature of $\sim 30$~$^{\circ}$C above their respective glass transition temperatures. 

\subsection{Film Transfer}
The polymer films were then cut into rectangular pieces and floated onto a pure water surface (MilliQ, Millipore inc). A slide coated with a high viscosity silicone oil (5100~Cp) and chilled to approximately -20~$^{\circ}$C, was gently pressed onto the floating film and removed from the bath taking the film with it. At this point, the film is floating on a silicone oil bath. The sample was allowed to float on the bath overnight in order to relax any stresses that were still present in the film from the floating and lifting process. Film thicknesses were measured by cutting pieces adjacent to the floated films and placing them on silicon wafers. Atomic force microscopy (AFM) was used to measure films of thickness 10~nm-1000~nm. Films thicker than this were imaged with a laser scanning confocal microscope.  Reflectance maxima (that corresponded to the top and bottom surfaces of the film) were extracted and were used with the index of refraction for the relevant polymer to find the thickness of each film.  Alternatively, the top surface of a film and the top surface of the substrate could be used near a sample edge.  No differences were noted between the various techniques.

A solution of glycerol (99.9\% Fisher Chemical) and fluorescent dye (fluorescein sodium, FUL-GLO) was created. A few drops of glycerol solution were deposited on the top of a film floating on (now room temperature) silicon oil, about 1 cm away from the oil-film edge.   To create a flat, elongated contact line, a glass slide with an edge parallel to both the film edge and the glycerol line, was placed in contact with the glycerol drops.  The geometry is shown in the schematics of Fig.~\ref{Expt}a,b.  The setup is allowed to equilibrate over  two hours, resulting in the state shown in Fig.~\ref{Expt}c.  The system was then scanned in three-dimensions using confocal microscopy (Olympus FLUOVIEW FV1000), from which the film and glycerol surfaces could be located (Fig.~\ref{Expt}d).

\begin{figure*}
 \centering
 \includegraphics[width = \textwidth]{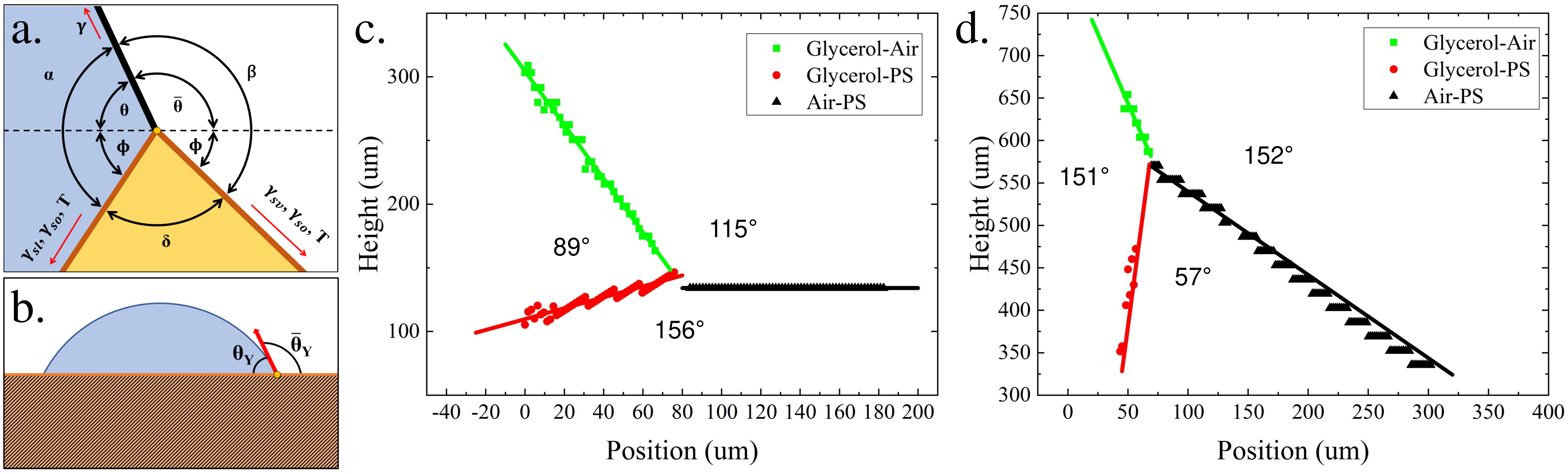}
 \caption{Surfaces around the triple line. a.) Schematic defining the various angles discussed in the text.  b.) The Young-Dupr\'{e} limit for reference.  c.) Experimental measurement of the triple line in a thick ($3.4$~$\mu$m) PS film.  Green represents the glycerol surface, black represents the film surface outside the fluid, and red represent the film surface below the glycerol.  Solid lines are linear fits used to determine the angles at the triple line  d.) Similar measurement with a much thinner film ($185$~nm).  The relative position of the polymer film has changed, while the fluid surface remains in a similar position. }
 \label{contactlines}
\end{figure*}

\subsection{Data Processing}
Intensity data was processed using ImageJ software and an algorithm to extract the peak intensity for each vertical slice in the three-dimensional scans. Reflectance and fluorescence channels were separated, allowing discrimination between the glycerol, glass slide, and polymer film surfaces. The height coordinate of submerged portions of films was then corrected to account for the refractive index of glycerol.  An example profile with the corrected film heights is given in Fig.~\ref{Expt}d. This profile data is used to find the height difference between the oil bath (far right of the film data) and the peak of the deformation. 

The relative angles of the glycerol-air interface, film in air, and film submerged in glycerol were extracted from the slopes of the data for these surfaces near the triple point.  Specifically, height data from $\sim$40 pixels away from the triple point along the glycerol surface ($\sim$65~$\mu$m) and from $\sim$75 pixel along the polymer surface ($\sim$120~$\mu$m) was used.  We refer to this length as the observation length for convenience, and denote it with the symbol $x_{obs}$.  The number of data points was selected in order to minimize distance from the triple line, while incorporating enough data points to account for noise in the measurements. Obvious out-lying data points were removed by hand (caused by intensity fluctuations for example).  Occasionally, the data this near the triple line was indeterminate when fitting an intensity peak. In this case, the data nearest to the triple line where an intensity peak could be reliably fit was used to determine the slope, with the error calculations being adjusted to reflect the shift away from the ideal measurement location.  The slopes of linear fits to the glycerol surface and the polymer surface data were used to calculate the relative angle between the glycerol and polymer surfaces.


\section{Results and Discussion}

A droplet of glycerol placed on top of a thin film will pull upwards and inwards on the films surface, deforming the film into the curved shape shown in figure~\ref{Expt}d.  The tension created by the glycerol surface is balanced at the triple line by the interplay of the outward pull of tension in the film, the change in surface energy of the film as the glycerol spreads, and the weight of the fluid displaced below the film.  These quantities are also intrinsically linked through bending in the plate as it deforms to accommodate the force balance.  If the film is thick, bending becomes the dominant energy, the plate remains flat, and the lift generated at the triple line is balanced by the weight of the fluid beneath the entire plate.  If the film is very thin, bending is less relevant, and the amount of fluid lifted is determined by the capillary length ($\sqrt{\gamma_{f}/\rho g}$, where $\gamma_{f}$ is the films surface energy, $\rho$ is the oil density, and $g$ the gravitational acceleration).  We note that in this geometry gravity and bending cannot be ignored in the thin film limit.  The main purpose of this paper is to clarify this complex interaction.

Figure~\ref{contactlines} shows the surfaces near the fluid/film contact line for a typical thick ($3.4$~$\mu$m) and thin ($185$~nm) PS film.  As it is of macroscopic convenience to discuss contact angles at a triple point, three angles at the contact line are directly measured, $\alpha$, $\beta$, and $\delta$.  We sub-divide $\alpha$ and $\beta$ into a portion above, $\theta$, and below, $\phi$, the horizontal for convenience (Fig.~\ref{contactlines}a).  Because we are focusing on small distances from the contact line, angle measurement can often be poorly defined because of the non-zero curvature near the film maxima.  Care must be taken in comparing an apparent angle with any theoretical predictions.  We extract angles from theory (discussed below) in the same way as we approach experimental data in order to maintain consistency.

In a thick film (Fig~\ref{contactlines}c.), the angle $\beta$ approaches a value of $115 ^{\circ}$, which is very close to what we measure independently with a sessile glycerol droplet on solid PS film (a PS film of $\sim 200$ nm which has been spin-coated on an atomically flat silicon substrate, see ESI\dag).  Contact with a thick film can be treated exactly as in the bulk; the contact angle is a result of a horizontal force balance which is independent of any tension in the film caused by external means.  In short, the film shows the classical balance of the Young-Dupr\'{e} law, $\gamma_{sl} + \gamma \cos(\theta_{Y})=\gamma_{sv}$, where $\gamma_{sl}$ and $\gamma_{sv}$ are the surface energies under the fluid or exposed to air respectively and $\gamma$ is the glycerol surface tension.  

As the film thins, bending becomes less important and the contact line lifts the film which results in the angles $\alpha$ and $\beta$ growing  (Fig.~\ref{contactlines}d).  To clarify, $\alpha$ is not equal to $\theta_{Y}$, and the air film interface is not horizontal as is often the case with higher tension free-standing films.\cite{Nadermann2013,Hui2014,Schulman2015,Schulman2017}  The glycerol surface, however, remains at a fairly constant angle with respect to the horizontal that is consistent with $\theta_{Y}$ determined from traditional sessile drop experiments ($\theta = 71 \pm 7 ^\circ$ for PS and $\theta = 69 \pm 9 ^\circ$ for PC measured in our experiment).  This suggests that the fluid shape is determined by the Young-Dupr\'{e} balance on a scale much smaller than the local radius of curvature where the thin polymer film still appears flat. 

This hypothesis also explains an apparent paradox that arises when considering a horizontal force balance, $\gamma \cos(\theta) + (T+\gamma_{sl}+\gamma_{so}) \cos(\phi)=(T+\gamma_{sv}+\gamma_{so}) \cos(\phi)$.  $\gamma_{so}$ refers to the oil film interface, $T$ to the tension acting on the ends of the plate, and the other variables remain as defined above.  The balance reduces to:
\begin{equation}\label{forceissue}
\cos(\theta)=\cos(\theta_{Y})\cos{\phi},
\end{equation}
and is satisfied only if $\phi = 0$ when $\theta=\theta_{Y}$.  This conflicts with the measured values of $\phi$ which are increasingly non-zero as the film thins.  The conclusion must be that the observed values of $\phi$ do not coincide with the direction of forces acting at the triple line in this range of film thicknesses.  The only other possibility is that the observed contact angle $\theta$ is not equal to the thermodynamic contact angle (the solid film limiting value of $\theta_{Y}$).  This latter point is inconsistent with our measurements of $\theta$ for most films, although the measurement error does not completely exclude small changes in the angle (particularly in the thinnest films).  Contact angles, measured when curvature is present, carry information about the film shape on the observed lenghtscale $x_{obs}$, not necessarily information about the local force balance.

\begin{figure}
 \centering
 \includegraphics[width = 0.5 \textwidth]{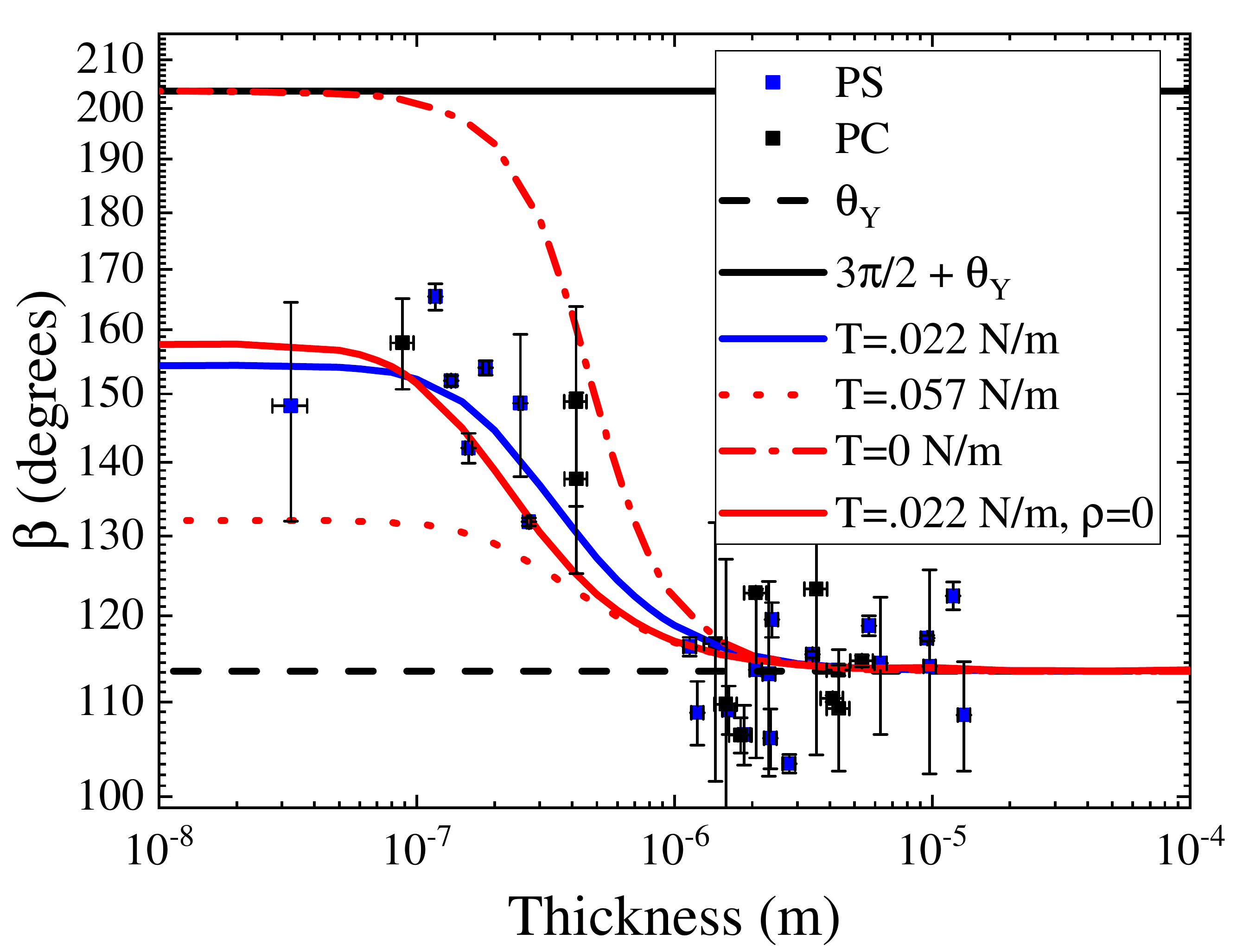}
 \caption{The external angle $\beta$ as a function of film thickness.  Data for PS films (blue squares) and PC (black circles) are shown along with the output of the numerical model (red curves).  The self-contact limit (solid black line) and the solid film (Young) limit (dashed black line) are also shown.}
 \label{angles}
\end{figure}

Figure~\ref{angles} shows the smooth change in $\beta$ as a function of thickness for both PS and PC films.  That both materials show such good overlap is consistent with their similar surface energies and material properties, and is also a sign that films are undamaged during processing and over the course of the experiment.  There are two independent physical limits for $\beta$.  First, $\beta$ must approach $\bar{\theta}_{Y} = \pi - \theta_{Y}$ at large thicknesses where the film does not bend.  Second, in the absence of bending and external tension acting on the free end of the film, the film will come into self contact and $\beta = \pi/2 + \bar{\theta}_{Y}$.  Measurements seem to indicate that this second limit is never reached experimentally.  As discussed above, a simple force balance is insufficient to describe how $\beta$ relates to surface forces, bending, and applied tension.  A fact made more clear when considering that no external forces or surface energies are related to thickness changes in our experiment.  A holistic approach which correctly accounts for bending moments, external tension and substrate density is required to describe observed contact angles.   

Treating the glycerol surface as a vertical line force of magnitude $\gamma \sin{\theta_{Y}}$, the system can be modeled as a Euler-Bernoulli beam on a Winkler foundation.  The foundation stiffness is determined by the density of the oil bath, $\rho$ and an external tension is supplied to the plate by the surface tension of the oil, $T$, pulling on the free end of the film.  The result is the fourth order differential equation:
\begin{equation}\label{Plate}
E I h^{\prime \prime \prime \prime}(x) = \gamma \delta(x) +T h^{\prime \prime}(x) + \rho g h(x),
\end{equation}
where $E$ is the Young's modulus of the plate, and $I=t^3/12(1-\nu^2)$ is the second moment of inertia per unit width.

Scaling analysis of Eqn.~\ref{Plate} reveals two important lengthscales.  If the tension term is much larger than the gravitational term, we find the problem to be scaled by a length of $x_{T}\sim \sqrt{Et^3/T}$ (as has been pointed out for free-standing films).\cite{Hui2014,Schulman2015}  In the opposite limit, we find a gravitational length of $x_{G} \sim (E t^3 /\rho g)^{1/4}$ to dominate the problem.\cite{Cerda2003}  Comparison of the two lengthscales defines a cross-over thickness, $t_{C}\sim (T^{2}/E\rho g)^{1/3}$, where the behaviour changes from gravity to tension dominated.  Using the plane strain modulus of polystyrene $E=3.3\times10^{8}$~Pa, the density and surface tension of the oil ($1110$~kg/m and 22~mN/m respectively) we predict a transition thickness of $t_{C}\sim 5 \mu$m, which matches the thickness at which $\beta$ begins to change from the solid film limit observed experimentally.    

If bending is ignored in Eqn.~\ref{Plate}, a capillary length, $x_{C}\sim\sqrt{T/\rho g}$, emerges from a similar scaling analysis.  Remarkably, the capillary length is equal to the gravitational and the tension lengths at $t_{c}$, highlighting the interdependence of externally applied tension, film bending, and gravity in the problem.  Nevertheless, the contact angles are determined by the smallest observable lengthscale, which in this case is $x_{T}$ when $t<t_{c}$.  Bending cannot be ignored at experimentally relevant thicknesses ($x_{T}$ is still greater than a micron for a $50$~nm thick film).  Even if the film were free of external tension, $x_{C} > x_{G}$ and is still not dominant at the contact line.

\begin{figure}
 \centering
 \includegraphics[width = 0.5 \textwidth]{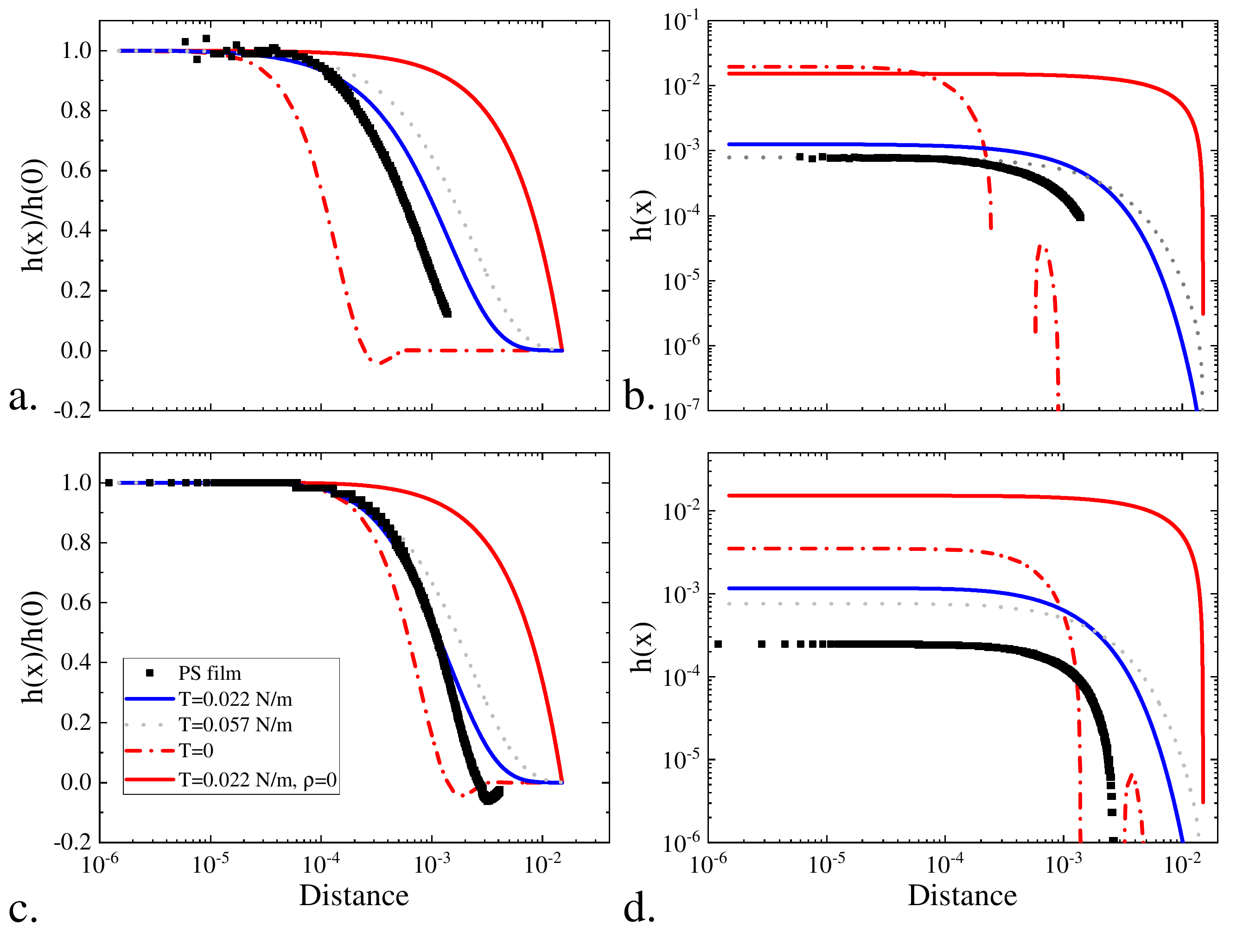}
 \caption{height as a function of thickness for $\sim 100$~nm (a,b) or $\sim 1$~$\mu$m (c,d) films.  Figures a. and c. show a linear log plot in which height has been normalized by its maximum value, in order for all curves to be visible. Figures b. and d. show the unnormalized height on log-log axis, again to facilitate viewing of all curves.  a. and b. show data from a $200$~nm PS film (black squares), and c. and d. show data from a $1.5$~$\mu$m film (black squares again).  Data shows good agreement even though none are `fit' by the model.}
 \label{shapes}
\end{figure}

\begin{figure}
 \centering
 \includegraphics[width = 0.5 \textwidth]{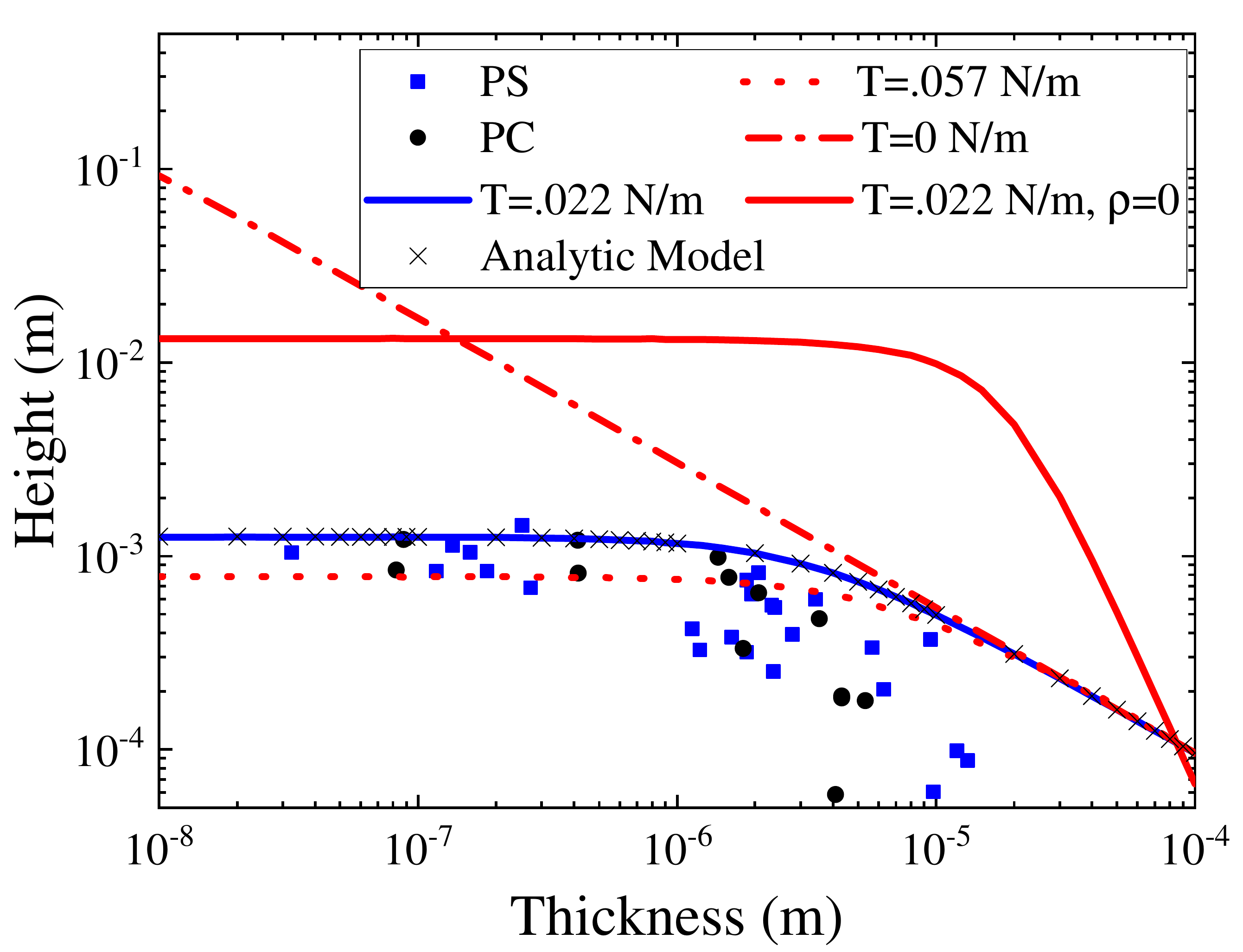}
 \caption{Maximum height of the triple line as a function of film thickness for PS (blue squares) and PC (black circles) films.  Numerical model results are shown for zero gravity, zero tension, and two other possible external tensions.  Analytic model results are shown as X's.}
 \label{height}
\end{figure}

To gain further insight, we solve Eqn.~\ref{Plate} numerically as a boundary value problem, and compare the numerical result with the experiment.  Sample film profiles of typical thin and thick films are shown in figure~\ref{shapes} a,b and c,d respectively.  Figure~\ref{angles} shows the predicted $\beta$ as a function of film thickness, where the angle was ``measured'' from numerically calculated shape profiles.  Figure~\ref{height} shows the peak height of the deformation as a function of film thickness alongside experimental measurements.  Several different sets of initial conditions were calculated, and four relevant results are shown in the figures.  

First, we calculate the zero gravity limit which eases comparison with existing free-standing film experiments.  The gravity free curve, assuming the external tension is still supplied by the oil phase surface tension, does fit the angle data but is clearly not physically related to our experiment as can be seen in its profile (Fig.~\ref{shapes}).  When $x_{obs} \sim x_{T}$ the angular data smoothly changes from solid like to a second limiting value as film thickness decreases.  The second transition occurs because $x_{obs} > x_{T}$, meaning the observation is not of high enough resolution to be influenced by $x_{T}$, it is now determined by a second lengthscale (in this case, the plate size $L$).  The shape taken on by the film in the absence of a fluid substrate is nearly triangular; flat with an upward slope far from the contact line, curving only at a lengthscale comparable to $x_{T}$.  The peak height is quite high (especially for thin films), and is strongly related to overall film length and the applied tension.  The discrepancy between the measured and simulated peak heights is another clear sign that gravity plays a role in our experiment, and a zero gravity approximation is not applicable.

Next we consider the contact line in the absence of an external tension (often the limit in which surface energies are measured\cite{Hui2014,Nadermann2013,Schulman2015}).  Here we find $\beta$ is considerably overestimated in comparison with the data, especially in the thin film limit where the film has come into self contact.  In this case, when $x_{obs}$ falls below $x_{T}$, the film shape is govern by the gravitational lengthscale, $x_{G}$, resulting in an increased drive to narrow the peak width.  At larger thicknesses the angle approaches $\bar{\theta} _{Y}$ as is expected.  The height of the lifted region of film follows a single power law, again deviating from the data considerably in the thin film limit.  The power law can easily be explained by using the scaling length derived above, $x_{G}$, in combination with Eqn.~\ref{Plate} to derive a natural height $h_{G}\sim \gamma (\rho g)^{-3/4} (EI)^{-1/4}$.  The agreement with the scaling clarifies how gravity plays a critical role over the entire range of film thicknesses in the zero tension limit.

Finally we show a curve generated using externally measured values for $\gamma$, $T$ and $\rho$.  This curve fits both the measured $\beta$ and also fits the height data quite well, with no additional free parameters.  While not inconsistent with a low thickness $\beta$ plateau, the data does not show a clear transition to `thin'.  This is likely due to the low brightness levels and tiny peak widths increasing error in our experiments coupled with the fragile nature of extremely thin films (fewer experiments survive processing).  The height data is much more reliable in the thin film limit, but shows some deviation in the thick film limit (where $\beta$ shows good agreement with the model).  This is due to a combination of experimental effects, including plate lengths being comparable to the contact region width, and the weight of the fluid resting on the film altering the boundary conditions (we use symmetric hinge boundaries and large plates in the model).  We additionally show model predictions for a film with a slightly larger tension to highlight the sensitivity of the experiment.  In this case the low thickness $\beta$ plateau occurs at a lower angle which is inconsistent with the data.

We supplement the beam model of Eqn.~\ref{Plate} by calculating similar height profiles from a free-energy based capillary model (see Appendix~A).  This method has the advantage of producing simple, analytic results for various measured properties and removes some of the complexity of a full continuum theory.  For the sake of clarity, we only focus on the maximum height of the deformation and contact angle in this work.  Assuming a zero derivative boundary at the deformation peak (consistent with our interpretation of the true contact point), we find the peak height as a function of thickness to be:
\begin{equation}\label{analyticheight}
h_{0}= x_{C} \frac{\gamma}{2 \gamma_{s}}\frac{1}{\sqrt{1+2(x_{G}^2/x_{C}^{2})}}. 
\end{equation}
where $\gamma_{s}$ is the surface energy associated with moving the plate upwards and increasing the surface length.  Because the plate's extension is tiny, we ignore any change in plate length and associate $\gamma_{s}$ with $T$ the surface tension of the oil bath.  Eqn.~\ref{analyticheight} is plotted in figure~\ref{height} alongside earlier numerical results, and the agreement is near perfect.  This shows that there is no difference between energy-based or force-based models, and again highlights the interdependence of gravity, tension and bending in our experiment.  

The contact angle in this model is simply $\theta_{Y}$ by construction (e.g. the zero derivative boundary condition at the deformation peak).  However, as experimental angles are measured a distance away from the true contact line ($x_{obs}$), the true contact angle is is not observed directly.  Again, as with the numerical model, we can derive an apparent angle from the predicted curve shape.  In this case, we can proceed analytically by using the derivative of the film shape (shown in the appendix) calculated at $x_{obs}$.  Figure~\ref{analyticangles} shows the resulting $\beta$ plotted alongside the PC and PS data with the numerical `fit' to ease comparison.  The analytic result is in good agreement with the numerical model and what is observed in the experiment.  The two result do not prove that the true contact angle is always $\theta_{Y}$, but shows experiments are at least consistent with this hypothesis.

\begin{figure}
 \centering
 \includegraphics[width = 0.5 \textwidth]{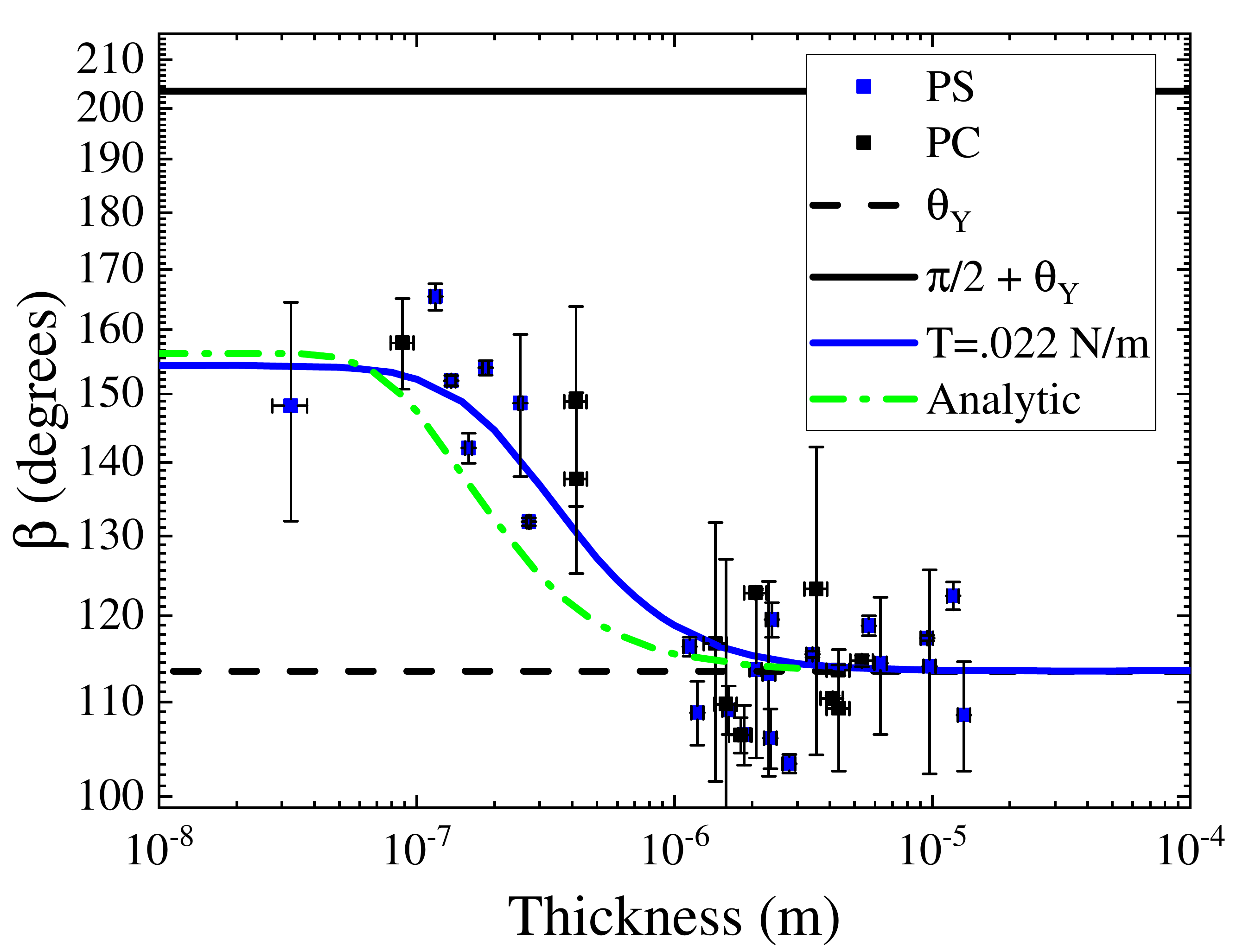}
 \caption{The external angle $\beta$ as a function of film thickness reproduced to clarify demonstration of the analytic theory discussed in the text.  Data for PS films (blue squares) and PC (black circles) are shown along with the output of the optimal numerical model (blue curve).  Analytic theory calculated with a $5 \mu$m observation length is shown by the dash-dotted green curve.}
 \label{analyticangles}
\end{figure}

\section{Conclusions}
We have examined the microscopic details of a fluid/thin-film contact line using confocal microscopy.  We find that gravity, external tension, and bending are all important in the region of film thicknesses examined ($\sim 1\times10^{-8}$ to $\sim 1\times10^{-5}$~m).  The experimentally measured angles, film shapes and peak heights show good agreement with with a Euler-Bernoulli beam or an equivalent, analytic, capillary model.  Our results show that a tension only force balance, as in the Neumann construction, is not possible over most of the range of our experiments.  Films with thicknesses above 100~nm have too much bending to consistently satisfy such a model, largely because the concept of a contact angle is ill defined.  We suggest that, a Young-Dupr{\'e} force balance always takes place on an extremely local scale, although may not be observable if the radius of curvature falls below optically observable lengthscales.  As film thicknesses increase, so does the radius of curvature, which leads to imprecise contact angles which are not useful in determining a force balance.  Eventually, bending completely dominates, the film remains globally flat and a Young-Dupr\'{e} horizontal force balance is macroscopically observable.

\section{Acknowledgements}
The authors gratefully acknowledge support from the AFOSR under the Young Investigator Program (FA9550-15-1-0168).

\section{Appendix A: Capillary Model}
A capillary model can be constructed to describe the free energy of a fluid resting on a thin film in one dimension.  Minimizing the difference in free-energy per unit width before and after the film is deformed allows analytic solutions to be developed in several useful situations.  From this point of view, each surface contributes a term to the free energy which is proportional to its surface energy, and its surface area, $A$.  For example, the fluid resting on top of the film has a free energy, $\int \gamma dA$ which can be reduced if the film is lifted and decreases the fluid/air interfacial area.  The fluid bath beneath the film is lifted if the film is deformed above the horizontal plane and thus contributes a term proportional to the height of the deformation and the density, $\rho$, of the fluid bath ($\rho g \int dx \int h(x) dh$).  The bending of the film as it is lifted also adds energy per unit width of film, $U_{B}=B/2\int \kappa ^{2} dx$ where $\kappa$ is the films curvature, $B=E t^3 /12(1-\nu^2)$ is the bending modulus with thickness $t$ and Poisson ratio $\nu$.  Assuming, for simplicity, the problem is symmetric about the origin and all deformations are small, the total change in free energy per unit width can be written:
\begin{equation}\label{freeE}
\Delta F = - \gamma h_{0} + \gamma_{S} \int _{0} ^{\infty}  h^{\prime 2} dx+ \rho g \int _{0} ^{\infty} h^{2} dx+ B \int _{0} ^{\infty} h^{\prime \prime 2} dx,
\end{equation}
where $h_{0}$ is the height at the origin (we assume a purely vertical fluid surface lifting the film), $\gamma$ is the top fluid surface tension, and $\gamma_{S}$ is the net surface energy change of the film and oil interface.  Using the capillary length, $x_{C}=\sqrt{\gamma _{S}/\rho g}$ and the gravity length $x_{G} = (EI/\rho g b)^{1/4}$, Eqn.~\ref{freeE} can be simplified to:  
\begin{equation}\label{freeE2}
\Delta F = - \gamma h_{0} + \rho g \int_{0} ^{\infty} \Bigg{[} x_{C}^{2} h^{\prime 2} + h^{2} + x_{G}^4  h^{\prime \prime 2} \Bigg{]}dx.
\end{equation}
Eqn.~\ref{freeE2} is minimized when the differential equation,
\begin{equation}\label{SylvioDE}
h-x_{c}^2 h^{\prime \prime} + x_{G}^{4} h^{\prime \prime \prime \prime} = 0,
\end{equation}
is satisfied.  In this case, the solution is given by
\begin{equation}\label{solution}
h(x)=\frac{h_{0}}{w_{1}-w_{2}}\bigg{(}w_{1}e^{-w_{2}x} -w_{2}e^{-w_{1}x} \bigg{)},
\end{equation}
where $w_1$ and $w_2$ can be created from the two rival lengthscales if $w_1^{2}w_{2}^{2}= 1/x_{G}^{4}$ and $w_1^{2}+w_{2}^{2}= 1/x_{C}$.  With this analytic solution and a set of boundary conditions, many useful properties can be calculated directly.  Most relevant to the current work, is the height at the origin, $h_{0}$, given by,
\begin{equation}
h_{0}= x_{C} \frac{\gamma}{2 \gamma_{s}}\frac{1}{\sqrt{1+2(x_{G}^2/x_{C}^{2})}}.
\end{equation}
A second useful expression, the slope of the curve, allows a determination of any apparent contact angle of interest.  The slope of Eqn.~\ref{solution} is simply its derivative, given by
\begin{equation}\label{derivative}
h^{\prime}(x)=\frac{h_{0}}{w_{1}-w_{2}}w_{1}w_{2}\bigg{(}-e^{-w_{2}x} +e^{-w_{1}x} \bigg{)}.
\end{equation}
For example, we use Eqn.~\ref{derivative} to calculate a contact angle for films of thickness ranging from $3 \mu$m to $10 $nm at a point $x=5 \mu$m, which is shown in figure~\ref{analyticangles}.



\balance


\bibliography{1DC} 

\providecommand*{\mcitethebibliography}{\thebibliography}
\csname @ifundefined\endcsname{endmcitethebibliography}
{\let\endmcitethebibliography\endthebibliography}{}
\begin{mcitethebibliography}{29}
\providecommand*{\natexlab}[1]{#1}
\providecommand*{\mciteSetBstSublistMode}[1]{}
\providecommand*{\mciteSetBstMaxWidthForm}[2]{}
\providecommand*{\mciteBstWouldAddEndPuncttrue}
  {\def\EndOfBibitem{\unskip.}}
\providecommand*{\mciteBstWouldAddEndPunctfalse}
  {\let\EndOfBibitem\relax}
\providecommand*{\mciteSetBstMidEndSepPunct}[3]{}
\providecommand*{\mciteSetBstSublistLabelBeginEnd}[3]{}
\providecommand*{\EndOfBibitem}{}
\mciteSetBstSublistMode{f}
\mciteSetBstMaxWidthForm{subitem}
{(\emph{\alph{mcitesubitemcount}})}
\mciteSetBstSublistLabelBeginEnd{\mcitemaxwidthsubitemform\space}
{\relax}{\relax}

\bibitem[Terfort \emph{et~al.}(1997)Terfort, Bowden, and
  Whitesides]{Terfort1997}
A.~Terfort, N.~Bowden and G.~Whitesides, \emph{Nature}, 1997, \textbf{386},
  162--164\relax
\mciteBstWouldAddEndPuncttrue
\mciteSetBstMidEndSepPunct{\mcitedefaultmidpunct}
{\mcitedefaultendpunct}{\mcitedefaultseppunct}\relax
\EndOfBibitem
\bibitem[Py \emph{et~al.}(2007)Py, Reverdy, Doppler, Bico, Roman, and
  Baroud]{py2007}
C.~Py, P.~Reverdy, L.~Doppler, J.~Bico, B.~Roman and C.~N. Baroud, \emph{Phys.
  Rev. Lett.}, 2007, \textbf{98}, 156103\relax
\mciteBstWouldAddEndPuncttrue
\mciteSetBstMidEndSepPunct{\mcitedefaultmidpunct}
{\mcitedefaultendpunct}{\mcitedefaultseppunct}\relax
\EndOfBibitem
\bibitem[Py \emph{et~al.}(2009)Py, Reverdy, Doppler, Bico, Roman, and
  Baroud]{py2009}
C.~Py, P.~Reverdy, L.~Doppler, J.~Bico, B.~Roman and C.~Baroud, \emph{Euro.
  Phys. J. Spec. Top.}, 2009, \textbf{166}, 67--71\relax
\mciteBstWouldAddEndPuncttrue
\mciteSetBstMidEndSepPunct{\mcitedefaultmidpunct}
{\mcitedefaultendpunct}{\mcitedefaultseppunct}\relax
\EndOfBibitem
\bibitem[Pericet-C{\'a}mara \emph{et~al.}(2008)Pericet-C{\'a}mara, Best, Butt,
  and Bonaccurso]{Pericet2008}
R.~Pericet-C{\'a}mara, A.~Best, H.-J. Butt and E.~Bonaccurso, \emph{Langmuir},
  2008, \textbf{24}, 10565--10568\relax
\mciteBstWouldAddEndPuncttrue
\mciteSetBstMidEndSepPunct{\mcitedefaultmidpunct}
{\mcitedefaultendpunct}{\mcitedefaultseppunct}\relax
\EndOfBibitem
\bibitem[Shanahan and Carre(1995)]{Shanahan1995}
M.~Shanahan and A.~Carre, \emph{Langmuir}, 1995, \textbf{11}, 1396--1402\relax
\mciteBstWouldAddEndPuncttrue
\mciteSetBstMidEndSepPunct{\mcitedefaultmidpunct}
{\mcitedefaultendpunct}{\mcitedefaultseppunct}\relax
\EndOfBibitem
\bibitem[Extrand and Kumagai(1996)]{Extrand1996}
C.~Extrand and Y.~Kumagai, \emph{Journal of colloid and interface science},
  1996, \textbf{184}, 191--200\relax
\mciteBstWouldAddEndPuncttrue
\mciteSetBstMidEndSepPunct{\mcitedefaultmidpunct}
{\mcitedefaultendpunct}{\mcitedefaultseppunct}\relax
\EndOfBibitem
\bibitem[Jerison \emph{et~al.}(2011)Jerison, Xu, Wilen, and
  Dufresne]{Jerison2011}
E.~R. Jerison, Y.~Xu, L.~A. Wilen and E.~R. Dufresne, \emph{Phys. Rev. Lett.},
  2011, \textbf{106}, 186103\relax
\mciteBstWouldAddEndPuncttrue
\mciteSetBstMidEndSepPunct{\mcitedefaultmidpunct}
{\mcitedefaultendpunct}{\mcitedefaultseppunct}\relax
\EndOfBibitem
\bibitem[Style and Dufresne(2012)]{Style2012}
R.~W. Style and E.~R. Dufresne, \emph{Soft Matter}, 2012, \textbf{8},
  7177--7184\relax
\mciteBstWouldAddEndPuncttrue
\mciteSetBstMidEndSepPunct{\mcitedefaultmidpunct}
{\mcitedefaultendpunct}{\mcitedefaultseppunct}\relax
\EndOfBibitem
\bibitem[Style \emph{et~al.}(2013)Style, Boltyanskiy, Che, Wettlaufer, Wilen,
  and Dufresne]{Style2013b}
R.~W. Style, R.~Boltyanskiy, Y.~Che, J.~Wettlaufer, L.~A. Wilen and E.~R.
  Dufresne, \emph{Phys. Rev. Lett.}, 2013, \textbf{110}, 066103\relax
\mciteBstWouldAddEndPuncttrue
\mciteSetBstMidEndSepPunct{\mcitedefaultmidpunct}
{\mcitedefaultendpunct}{\mcitedefaultseppunct}\relax
\EndOfBibitem
\bibitem[Xu \emph{et~al.}(2018)Xu, Style, and Dufresne]{Qin2018}
Q.~Xu, R.~W. Style and E.~R. Dufresne, \emph{Soft Matter}, 2018, \textbf{14},
  916--920\relax
\mciteBstWouldAddEndPuncttrue
\mciteSetBstMidEndSepPunct{\mcitedefaultmidpunct}
{\mcitedefaultendpunct}{\mcitedefaultseppunct}\relax
\EndOfBibitem
\bibitem[Pham \emph{et~al.}(2017)Pham, Schellenberger, Kappl, and
  Butt]{Pham2017}
J.~T. Pham, F.~Schellenberger, M.~Kappl and H.-J. Butt, \emph{Phys. Rev. Mat.},
  2017, \textbf{1}, 015602\relax
\mciteBstWouldAddEndPuncttrue
\mciteSetBstMidEndSepPunct{\mcitedefaultmidpunct}
{\mcitedefaultendpunct}{\mcitedefaultseppunct}\relax
\EndOfBibitem
\bibitem[Style \emph{et~al.}(2013)Style, Hyland, Boltyanskiy, Wettlaufer, and
  Dufresne]{Style2013a}
R.~W. Style, C.~Hyland, R.~Boltyanskiy, J.~S. Wettlaufer and E.~R. Dufresne,
  \emph{Nat. Comm.}, 2013, \textbf{4}, 2728\relax
\mciteBstWouldAddEndPuncttrue
\mciteSetBstMidEndSepPunct{\mcitedefaultmidpunct}
{\mcitedefaultendpunct}{\mcitedefaultseppunct}\relax
\EndOfBibitem
\bibitem[Hui \emph{et~al.}(2002)Hui, Jagota, Lin, and Kramer]{Hui2002}
C.~Y. Hui, A.~Jagota, Y.~Y. Lin and E.~J. Kramer, \emph{Langmuir}, 2002,
  \textbf{18}, 1394--1407\relax
\mciteBstWouldAddEndPuncttrue
\mciteSetBstMidEndSepPunct{\mcitedefaultmidpunct}
{\mcitedefaultendpunct}{\mcitedefaultseppunct}\relax
\EndOfBibitem
\bibitem[Mora \emph{et~al.}(2010)Mora, Phou, Fromental, Pismen, and
  Pomeau]{Mora2010}
S.~Mora, T.~Phou, J.-M. Fromental, L.~M. Pismen and Y.~Pomeau, \emph{Phys. Rev.
  Lett.}, 2010, \textbf{105}, 214301\relax
\mciteBstWouldAddEndPuncttrue
\mciteSetBstMidEndSepPunct{\mcitedefaultmidpunct}
{\mcitedefaultendpunct}{\mcitedefaultseppunct}\relax
\EndOfBibitem
\bibitem[Andreotti \emph{et~al.}(2016)Andreotti, B{\"a}umchen, Boulogne,
  Daniels, Dufresne, Perrin, Salez, Snoeijer, and Style]{Andreotti2016}
B.~Andreotti, O.~B{\"a}umchen, F.~Boulogne, K.~E. Daniels, E.~R. Dufresne,
  H.~Perrin, T.~Salez, J.~H. Snoeijer and R.~W. Style, \emph{Soft Matter},
  2016, \textbf{12}, 2993--2996\relax
\mciteBstWouldAddEndPuncttrue
\mciteSetBstMidEndSepPunct{\mcitedefaultmidpunct}
{\mcitedefaultendpunct}{\mcitedefaultseppunct}\relax
\EndOfBibitem
\bibitem[Style \emph{et~al.}(2017)Style, Jagota, Hui, and Dufresne]{Style2017}
R.~W. Style, A.~Jagota, C.-Y. Hui and E.~R. Dufresne, \emph{Annu. Rev. of Cond.
  Matt. Phys.}, 2017, \textbf{8}, 99--118\relax
\mciteBstWouldAddEndPuncttrue
\mciteSetBstMidEndSepPunct{\mcitedefaultmidpunct}
{\mcitedefaultendpunct}{\mcitedefaultseppunct}\relax
\EndOfBibitem
\bibitem[Bae \emph{et~al.}(2015)Bae, Ouchi, and Hayward]{Bae2015}
J.~Bae, T.~Ouchi and R.~C. Hayward, \emph{ACS Appl. Mat. Int.}, 2015,
  \textbf{7}, 14734--14742\relax
\mciteBstWouldAddEndPuncttrue
\mciteSetBstMidEndSepPunct{\mcitedefaultmidpunct}
{\mcitedefaultendpunct}{\mcitedefaultseppunct}\relax
\EndOfBibitem
\bibitem[Huang \emph{et~al.}(2007)Huang, Juszkiewicz, De~Jeu, Cerda, Emrick,
  Menon, and Russell]{Huang2007}
J.~Huang, M.~Juszkiewicz, W.~H. De~Jeu, E.~Cerda, T.~Emrick, N.~Menon and T.~P.
  Russell, \emph{Science}, 2007, \textbf{317}, 650--653\relax
\mciteBstWouldAddEndPuncttrue
\mciteSetBstMidEndSepPunct{\mcitedefaultmidpunct}
{\mcitedefaultendpunct}{\mcitedefaultseppunct}\relax
\EndOfBibitem
\bibitem[Schroll \emph{et~al.}(2013)Schroll, Adda-Bedia, Cerda, Huang, Menon,
  Russell, Toga, Vella, and Davidovitch]{Schroll2013}
R.~Schroll, M.~Adda-Bedia, E.~Cerda, J.~Huang, N.~Menon, T.~Russell, K.~Toga,
  D.~Vella and B.~Davidovitch, \emph{Phys. Rev. Lett.}, 2013, \textbf{111},
  014301\relax
\mciteBstWouldAddEndPuncttrue
\mciteSetBstMidEndSepPunct{\mcitedefaultmidpunct}
{\mcitedefaultendpunct}{\mcitedefaultseppunct}\relax
\EndOfBibitem
\bibitem[Paulsen \emph{et~al.}(2016)Paulsen, Hohlfeld, King, Huang, Qiu,
  Russell, Menon, Vella, and Davidovitch]{Paulsen2016}
J.~D. Paulsen, E.~Hohlfeld, H.~King, J.~Huang, Z.~Qiu, T.~P. Russell, N.~Menon,
  D.~Vella and B.~Davidovitch, \emph{Proc. Nat. Acad. Sci.}, 2016,
  \textbf{113}, 1144--1149\relax
\mciteBstWouldAddEndPuncttrue
\mciteSetBstMidEndSepPunct{\mcitedefaultmidpunct}
{\mcitedefaultendpunct}{\mcitedefaultseppunct}\relax
\EndOfBibitem
\bibitem[Paulsen \emph{et~al.}(2017)Paulsen, D{\'e}mery, Toga, Qiu, Russell,
  Davidovitch, and Menon]{Paulsen2017}
J.~D. Paulsen, V.~D{\'e}mery, K.~B. Toga, Z.~Qiu, T.~P. Russell, B.~Davidovitch
  and N.~Menon, \emph{Phys. Rev. Lett.}, 2017, \textbf{118}, 048004\relax
\mciteBstWouldAddEndPuncttrue
\mciteSetBstMidEndSepPunct{\mcitedefaultmidpunct}
{\mcitedefaultendpunct}{\mcitedefaultseppunct}\relax
\EndOfBibitem
\bibitem[Paulsen \emph{et~al.}(2015)Paulsen, D\'{e}mery, Santangelo, T.P.,
  Davidovitch, and Mennon]{Paulsen2015}
J.~D. Paulsen, V.~D\'{e}mery, C.~Santangelo, R.~T.P., B.~Davidovitch and
  N.~Mennon, \emph{Nat. Mat.}, 2015, \textbf{14}, 1206--1209\relax
\mciteBstWouldAddEndPuncttrue
\mciteSetBstMidEndSepPunct{\mcitedefaultmidpunct}
{\mcitedefaultendpunct}{\mcitedefaultseppunct}\relax
\EndOfBibitem
\bibitem[Kumar \emph{et~al.}(2018)Kumar, Paulsen, T.P., and Mennon]{Kumar2018}
D.~Kumar, J.~D. Paulsen, R.~T.P. and N.~Mennon, \emph{Science}, 2018,
  \textbf{359}, 775--778\relax
\mciteBstWouldAddEndPuncttrue
\mciteSetBstMidEndSepPunct{\mcitedefaultmidpunct}
{\mcitedefaultendpunct}{\mcitedefaultseppunct}\relax
\EndOfBibitem
\bibitem[Schulman and Dalnoki-Veress(2015)]{Schulman2015}
R.~D. Schulman and K.~Dalnoki-Veress, \emph{Phys. Rev. Lett.}, 2015,
  \textbf{115}, 206101\relax
\mciteBstWouldAddEndPuncttrue
\mciteSetBstMidEndSepPunct{\mcitedefaultmidpunct}
{\mcitedefaultendpunct}{\mcitedefaultseppunct}\relax
\EndOfBibitem
\bibitem[Schulman \emph{et~al.}(2015)Schulman, Ledesma-Alonso, Salez,
  Rapha\"{e}l, and Dalnoki-Veress]{Schulman2017}
R.~D. Schulman, R.~Ledesma-Alonso, T.~Salez, Rapha\"{e}l and K.~Dalnoki-Veress,
  \emph{Phys. Rev. Lett.}, 2015, \textbf{118}, 198002\relax
\mciteBstWouldAddEndPuncttrue
\mciteSetBstMidEndSepPunct{\mcitedefaultmidpunct}
{\mcitedefaultendpunct}{\mcitedefaultseppunct}\relax
\EndOfBibitem
\bibitem[Schulman \emph{et~al.}(2018)Schulman, Trejo, Salez, Rapha\"{e}l, and
  Dalnoki-Veress]{Schulman2018}
R.~D. Schulman, M.~Trejo, T.~Salez, Rapha\"{e}l and K.~Dalnoki-Veress,
  \emph{Nat. Comm.}, 2018, \textbf{9}, 982\relax
\mciteBstWouldAddEndPuncttrue
\mciteSetBstMidEndSepPunct{\mcitedefaultmidpunct}
{\mcitedefaultendpunct}{\mcitedefaultseppunct}\relax
\EndOfBibitem
\bibitem[Nadermann \emph{et~al.}(2013)Nadermann, Hui, and
  Jagota]{Nadermann2013}
N.~Nadermann, C.-Y. Hui and A.~Jagota, \emph{Proc. Nat. Acad. Sci.}, 2013,
  \textbf{110}, 10541--10545\relax
\mciteBstWouldAddEndPuncttrue
\mciteSetBstMidEndSepPunct{\mcitedefaultmidpunct}
{\mcitedefaultendpunct}{\mcitedefaultseppunct}\relax
\EndOfBibitem
\bibitem[Hui and Jagota(2014)]{Hui2014}
C.-Y. Hui and A.~Jagota, \emph{Proc. R. Soc. A}, 2014, \textbf{470},
  20140085\relax
\mciteBstWouldAddEndPuncttrue
\mciteSetBstMidEndSepPunct{\mcitedefaultmidpunct}
{\mcitedefaultendpunct}{\mcitedefaultseppunct}\relax
\EndOfBibitem
\bibitem[Cerda and Mahadevan(2003)]{Cerda2003}
E.~Cerda and L.~Mahadevan, \emph{Phys. Rev. Lett.}, 2003, \textbf{90},
  074302\relax
\mciteBstWouldAddEndPuncttrue
\mciteSetBstMidEndSepPunct{\mcitedefaultmidpunct}
{\mcitedefaultendpunct}{\mcitedefaultseppunct}\relax
\EndOfBibitem
\end{mcitethebibliography}
\bibliographystyle{rsc} 

\end{document}